\documentclass[11pt]{article}%
\usepackage{amsmath}
\usepackage{amsfonts}
\usepackage{amssymb}
\usepackage{graphicx}
\usepackage{braket}
\usepackage[english]{babel}
\usepackage[numbers,sort&compress]{natbib}
\usepackage[colorlinks=true,urlcolor=blue,citecolor=blue]{hyperref}
\usepackage{url}
\usepackage{doi}%
\providecommand{\U}[1]{\protect\rule{.1in}{.1in}}

\begin{document}

\title{On a Class of Time-Dependent Non-Hermitian Hamiltonians}
\author{F. Kecita$^{a,b}$\thanks{kecita.f@centre-univ-mila.dz}, \ B. Khantoul$^{a,c}%
$\thanks{boubakeur.khantoul@univ-constantine3.dz}, \ A. Bounames$^{a}%
$\thanks{E-mail: bounames@univ-jijel.dz \ }\\$^{(a)}$ Laboratory of Theoretical Physics, Department of Physics, \\University of Jijel, 18000 Jijel, Algeria.\\$^{(b)}$ Department of Process Engineering, Institute of Science and Technology,\\\ University Center of Mila, 43000 Mila, Algeria.\\$^{(c)}$Department of Process Engineering, University of Constantine 3 -- \\\ \ Salah Boubnider, 25000 Constantine, Algeria.}
\date{}
\maketitle

\begin{abstract}
We study a class of time-dependent (TD) non-Hermitian Hamiltonians $H(t)$ that can be transformed into a time-independent pseudo-Hermitian Hamiltonian $\mathcal{H}_{0}^{PH}$ using a suitable TD unitary transformation $F(t)$. The latter can in turn be related to a
Hermitian Hamiltonian $h$ by a similarity transformation, $h=\rho
\mathcal{H}_{0}^{PH} \rho^{-1}$ where $\rho$ is the Dyson map. Accordingly,
once the Schr\"{o}dinger equation for the Hermitian Hamiltonian $h$ is solved,
the general solution of the initial system can be deduced. This allows to
define the appropriate $\tilde{\eta}(t)$-inner product for the Hilbert space
associated with $H(t)$, where $\tilde{\eta}(t)=F^{\dagger}(t)\eta F(t)$ and
$\eta=\rho^{\dagger}\rho$ is the metric operator. This greatly simplifies the
computation of the relevant uncertainty relations for these systems. As an
example, we consider a model of a particle with a TD mass subjected to a specific TD
complex linear potential. We thus obtain two Hermitian Hamiltonians, namely that of the 
standard harmonic oscillator and that of the inverted oscillator. For both cases, the auxiliary equation admits a
solution, and the exact analytical solutions are squeezed states given in terms of the 
Hermite polynomials with complex coefficients. Moreover, when the Hermitian Hamiltonian is that of the harmonic oscillator, the position-momentum uncertainty relation is real and greater than or equal to $\hbar/2$, thereby confirming its consistency.
\newline\newline

Keywords: Non-Hermitian Hamiltonian, time-dependent Hamiltonian,
pseudo-Hermiticity, metric operator, uncertainty relation, unitary
transformation, similarity transformation.

\end{abstract}

\section{Introduction}

In quantum mechanics, observables are represented by Hermitian operators to
ensure real eigenvalues. The Hilbert space of state vectors is endowed with an
inner product and having a positive norm. The Hermiticity of the Hamiltonian
guarantees unitary time evolution and preserves the norm of the quantum state
over time, which means that probability is conserved. However, a significant
discovery has been made that Hamiltonians invariant under $\mathcal{PT}%
$-symmetry have real eigenvalues in the unbroken phase, despite being
non-Hermitian in the conventional sense \cite{bender1}. This result has
stimulated much research into these new quantum systems
\cite{znojil1,bender2,bender3}. The other important advance is the use of the
concept of pseudo-hermiticity. In this setting, a Hamiltonian $H$ is said to
be pseudo-Hermitian if there exists a linear, invertible and Hermitian metric
operator $\eta$ such that \cite{Dieud,scholtz}
\begin{equation}
H^{\dagger}=\eta H\eta^{-1},
\end{equation}
which guarantees that $H$ is similar to a Hermitian operator $h$, via the
following similarity transformation \cite{most1}
\begin{equation}
h=\rho H\rho^{-1},
\end{equation}
where the Dyson map $\rho$ satisfies $\eta=\rho^{\dagger}\rho$. As a
consequence, the spectrum of $H$ is thus real and its eigenstates
$|\phi\rangle$ are connected to those of $h$, namely $|\varphi\rangle$, through
the relationship $|\phi\rangle=\rho^{-1}|\varphi\rangle$ \cite{most1}.

Although these mathematical constructions have successfully extended the scope
of quantum mechanics to include non-Hermitian systems, they pose several
technical challenges, particularly how to define the appropriate inner product
in a consistent manner. For pseudo-Hermitian quantum systems, the standard
Dirac inner product is inadequate for ensuring the unitarity of time evolution
and the reality of observables. Instead, a modified inner product of the form
\cite{most1}
\begin{equation}
\langle\psi_{1}|\psi_{2}\rangle_{\eta}=\langle\psi_{1}|\eta|\psi_{2}\rangle
\end{equation}
is introduced to maintain the probabilistic interpretation of the theory.
However, constructing a suitable metric operator $\eta$ becomes especially
non-trivial particularly when the Hamiltonian is time-dependent.

On the other hand, the study of TD non-Hermitian quantum systems raises open
mathematical and conceptual questions
\cite{znojil08,znojil09,most5,fring2,znojil24}, and solving the associated
time-dependent Schr\"{o}dinger equation (TDSE) is still receiving increasing
attention
\cite{menouar,desousa,yuce1,faria1,most4,znojil2,wang1,fring1,khant1,maamache1,maamache2,ramos,pedrosa,ponte,
luiz1,koussa1,koussa2,comment,choi,elhar,dourado,zelaya,silva,cius,boug,amaouche,taira,khant2,maamache24,moussa1}%
. Standard approaches often require the construction of Dyson maps or metric
operators, which may be time-dependent, complicating both the interpretation
and the computation. Moreover, the emergence of a nonlinear Ermakov type
auxiliary equation, which is nontrivial to solve, constitutes another
constraint for obtaining exact analytical solutions \cite{32,33}. This
significantly reduces the number of exactly solvable TD non-Hermitian systems
\cite{34,35,36,38,kecita,cessa2}.

In this work, we use a unitary transformation $F(t)$ to explore a class of
non-Hermitian TD Hamiltonians $H(t)$ convertible into time-independent
pseudo-Hermitian Hamiltonians, $\mathcal{H}_{0}^{PH}$, itself related to a
Hermitian Hamiltonian $h$ by a similarity transformation, $h=\rho
\mathcal{H}_{0}^{PH} \rho^{-1}$ where $\rho$ is the Dyson map. We then solve
the Schr\"{o}dinger equation for the Hermitian Hamiltonian $h$, and deduce the
general solution of the initial system. This facilitates the definition of the
corresponding $\tilde{\eta}(t)$-inner product for the Hilbert space associated
with $H(t)$, where $\tilde{\eta}(t)=F^{\dagger}(t)\eta F(t)$ and $\eta
=\rho^{\dagger}\rho$ is the metric operator, and the uncertainty relations become
easy to compute.

This paper is organized as follows, in section 2, we present a method for solving the Schrodinger 
equation for a class of TD non-Hermitian Hamiltonians $H(t)$ using an appropriate TD unitary
transformation $F(t)$. Then, using the corresponding time-independent metric operator to define the appropriate
$\tilde{\eta}(t)$-inner product for the Hilbert space associated with $H(t)$.
In section 3, we derive the appropriate uncertainty relation for
pseudo-Hermitian observables $X$ and $P$ using the above $\tilde{\eta}%
(t)$-inner product. In section 4, as an application, we treat the model a
model of a particle with TD mass in a TD complex linear potential. At last, 
section 5 is devoted to the conclusion.

\section{Inner Product for TD Non-Hermitian Systems}

The time evolution of quantum systems, described by TD non-Hermitian Hamiltonians $H(t)$, is governed 
by the TDSE (with $\hbar=1$).
\begin{equation}
i\frac{\partial}{\partial t}|\psi(t)\rangle=H(t)|\psi(t)\rangle.
\label{schrod}%
\end{equation}
We introduce a time-dependent unitary transformation
$F(t)$, which maps the original state $|\psi(t)\rangle$ to a new state
$|\phi(t)\rangle$ via
\begin{equation}
|\phi(t)\rangle=F(t)|\psi(t)\rangle. \label{transf}%
\end{equation}

Substituting this into Eq.~(\ref{schrod}), we obtain a transformed
Schr\"{o}dinger equation of the form
\begin{equation}
i\frac{\partial}{\partial t}|\phi(t)\rangle=\mathcal{H}(t)|\phi(t)\rangle
,\label{modif}%
\end{equation}
where the effective Hamiltonian $\mathcal{H}(t)$ is given by
\begin{equation}
\mathcal{H}(t)=F(t)H(t)F^{\dagger}(t)-iF(t)\frac{\partial F^{\dagger}%
(t)}{\partial t}. \label{A3}%
\end{equation}

Suppose that under a suitable choice of $F(t)$, the Hamiltonian (\ref{A3})
becomes time-independent. The resulting Hamiltonian is pseudo-Hermitian,
denoted $\mathcal{H}_{0}^{PH}$, if there exists a positive definite metric
operator $\eta$ such that
\begin{equation}
\left( \mathcal{H}_{0}^{PH}\right) ^{\dagger}= \eta\mathcal{H}_{0}^{PH}
\eta^{-1},
\end{equation}
this leads to a generalized inner product defined as
\begin{equation}
\langle\phi(t)|\phi(t)\rangle_{\eta}=\langle\phi(t)|\eta|\phi(t)\rangle,
\end{equation}
which ensures unitary time evolution in the corresponding Hilbert space.

Since $\mathcal{H}_{0}^{PH}$ is time-independent, the solution of the equation
(\ref{modif}) can then be written as
\begin{equation}
|\phi(t)\rangle=e^{-iEt}|\phi_{0}\rangle,
\end{equation}
where $E$ and $|\phi_{0}\rangle$ denote the eigenvalue and eigenstate of the
time-independent Hamiltonian $\mathcal{H}_{0}^{PH}$, respectively.

Furthermore, the pseudo-Hermitian Hamiltonian $\mathcal{H}_{0}^{PH}$ can be
mapped to a Hermitian operator $h$ through a similarity transformation
\begin{equation}
h=\rho\mathcal{H}_{0}^{PH}\rho^{-1},\label{similarity}%
\end{equation}
where the metric operator is constructed as $\eta=\rho^{\dagger}\rho$. The
corresponding eigenstates satisfy
\begin{equation}
|\phi_{0}\rangle=\rho^{-1}|\varphi\rangle,
\end{equation}
with $|\varphi\rangle$ an eigenstate of the Hermitian Hamiltonian $h$.
Therefore, $\mathcal{H}_{0}^{PH}$ and $h$ are isospectral operators, that is,
they have the same spectrum E.

This framework allows us to define a consistent inner product for the TD
non-Hermitian system, as
\begin{equation}
\langle\phi(t)|\phi(t)\rangle_{\eta}=\langle\phi(t)|\rho^{\dagger}\rho
|\phi(t)\rangle=\langle\psi(t)|F^{\dagger}(t)\rho^{\dagger}\rho F(t)|\psi
(t)\rangle=\langle\psi(t)|\tilde{\eta}(t)|\psi(t)\rangle=\langle\psi
(t)|\psi(t)\rangle_{\tilde{\eta}(t)} ,
\end{equation}
where the TD metric operator associated with $H(t)$ takes the form
\begin{equation}
\tilde{\eta}(t)=F^{\dagger}(t)\rho^{\dagger}\rho F(t),
\end{equation}
and under this construction, the general solution of Eq. (\ref{schrod}) is
\begin{equation}
\lvert\psi(t)\rangle= F^{-1}(t)|\phi(t)\rangle=F^{-1}(t)\,e^{-\,iE t}\,\rho^{-1}\lvert\varphi\rangle, \label{gsol}
\end{equation}
where its norm, associated with the $\tilde{\eta}(t)$-inner product, is
preserved over time.

\section{Uncertainty Relations}

Heisenberg uncertainty relations were first developed for Hermitian operators
\cite{heisenb,robert}. Subsequent studies have extended uncertainty relations to non-Hermitian operators by endowing the Hilbert space, associated with each Hamiltonian, with an appropriate inner product \cite{13,14,fan2,garcia,zhao,shukla,bagarello,khant3}. In order to construct the expectation values and variances, we start from the general definition for a non-Hermitian operator $\mathcal{A}$ \cite{bagarello}, namely
\begin{equation}
\left(  \Delta\mathcal{A}\right)  ^{2}=\left\langle \mathcal{A}^{\dagger
}\mathcal{A}\right\rangle -\left\langle \mathcal{A}^{\dagger}\right\rangle
\left\langle \mathcal{A}\right\rangle ,
\end{equation}
and for a pseudo-Hermitian observable $\mathcal{A}$ with the $\tilde{\eta}%
(t)$-inner product such that $\mathcal{A}^{\dagger}=\tilde{\eta}%
\mathcal{A}\tilde{\eta}^{-1}$, we find that
\begin{equation}
\left\langle \mathcal{A}^{\dagger}\right\rangle _{\tilde{\eta}(t)}%
=\left\langle \psi(t)\right\vert \mathcal{A}^{\dagger}\tilde{\eta
}(t)\left\vert \psi(t)\right\rangle =\left\langle \psi(t)\right\vert
\tilde{\eta}(t)\mathcal{A}\left\vert \psi(t)\right\rangle =\left\langle
\mathcal{A}\right\rangle _{\tilde{\eta}(t)},
\end{equation}
\begin{equation}
\left\langle \mathcal{A}^{\dagger}\mathcal{A}\right\rangle _{\tilde{\eta}%
(t)}=\left\langle \psi(t)\right\vert \mathcal{A}^{\dagger}\tilde{\eta
}(t)\mathcal{A}\left\vert \psi(t)\right\rangle =\left\langle \psi
(t)\right\vert \tilde{\eta}(t)\mathcal{AA}\left\vert \psi(t) \right\rangle
=\left\langle \mathcal{A}^{2}\right\rangle _{\tilde{\eta}(t)}.
\end{equation}
Thus
\begin{equation}
\left(  \Delta\mathcal{A}\right)  _{\mathcal{\tilde{\eta}}(t)}^{2}%
=\left\langle \mathcal{A}^{2}\right\rangle _{\tilde{\eta}(t)}-\left\langle
\mathcal{A}\right\rangle _{\tilde{\eta}(t)}^{2}. \label{quad}
\end{equation}

In conventional quantum mechanics, the position $x$ and momentum $p$ operators
are Hermitian. However, within the pseudo-Hermitian framework, their
properties become dependent on the choice of the metric operator $\tilde{\eta
}(t)$. To derive the appropriate uncertainty relation in this context, we
introduce pseudo-Hermitian observables $X$ and $P$, constructed from $x$ and
$p$ in such a way that they remain physically consistent
\begin{equation}
	X\text{ }=\tilde{\rho}^{-1}(t)x\tilde{\rho}(t),\text{\ \ \ \ \ \ \ \ \ \ }%
	P=\tilde{\rho}^{-1}(t)p\tilde{\rho}(t),
\end{equation}
where $\tilde{\rho}(t)=\rho F(t)$, and $\tilde{\eta}(t)=\tilde{\rho}^{\dagger
}(t)\tilde{\rho}(t)$.

Following the methodology outlined in Refs. \cite{shukla,bagarello}, we apply
the Schwarz inequality with respect to the $\mathcal{\tilde{\eta}}(t)$-inner
product on $X$ and $P$ operators, then we get

\begin{equation}
\left(  \Delta X\right)  _{\mathcal{\tilde{\eta}}(t)}\left(  \Delta P\right)
_{\mathcal{\tilde{\eta}}(t)}=\left\Vert \tilde{X}\psi(t)\right\Vert
_{\mathcal{\tilde{\eta}}(t)}\left\Vert \tilde{P}\psi(t)\right\Vert
_{\mathcal{\tilde{\eta}}(t)}\geq\left\vert \left\langle \tilde{X}\tilde
{P}\right\rangle _{\mathcal{\tilde{\eta}}(t)}\right\vert, \label{uu1}%
\end{equation}
where
\begin{equation}
\left\vert \left\langle \tilde{X}\tilde{P}\right\rangle _{\mathcal{\tilde
{\eta}}(t)}\right\vert =\frac{1}{2}\left\vert \left\langle \left[  X,P\right]
\right\rangle _{\mathcal{\tilde{\eta}}(t)}+\left\langle \left\{  \tilde{X}%
^{+},\tilde{P}\right\}  \right\rangle _{\mathcal{\tilde{\eta}}(t)}\right\vert
\geq\frac{1}{2}\left\vert \left\langle \left[  X,P\right]  \right\rangle
_{\mathcal{\tilde{\eta}}(t)}\right\vert,
\end{equation}
and%
\begin{equation}
\tilde{X}=X-\left\langle X\right\rangle _{\mathcal{\tilde{\eta}}(t)},\text{
\ \ \ \ }\left\langle X\right\rangle _{\mathcal{\tilde{\eta}}(t)}=\left\langle
\psi(t),X\psi(t)\right\rangle _{\mathcal{\tilde{\eta}}(t)}.\text{\ \ \ }%
\end{equation}

Then, from Eq. (\ref{uu1}) we deduce that
\begin{equation}
\left(  \Delta X\right)  _{\mathcal{\tilde{\eta}}(t)}\left(  \Delta P\right)
_{\mathcal{\tilde{\eta}}(t)}\geq\frac{1}{2}\left\vert \left\langle \left[
X,P\right]  \right\rangle _{\mathcal{\tilde{\eta}}(t)}\right\vert ,
\label{C01}%
\end{equation}
and using Eq. (\ref{quad}), $\left(  \Delta X\right)_{\mathcal{\tilde{\eta}}(t)}$ and $\left(
\Delta P\right)_{\mathcal{\tilde{\eta}}(t)}$ are defined as
\begin{equation}
\left(  \Delta X\right)  _{\mathcal{\tilde{\eta}}(t)}^{2}=\left\langle
X^{2}\right\rangle _{\mathcal{\tilde{\eta}}(t)}-\left\langle X\right\rangle
_{\mathcal{\tilde{\eta}}(t)}^{2}, \label{C03}%
\end{equation}%
\begin{equation}
\left(  \Delta P\right)  _{\mathcal{\tilde{\eta}}(t)}^{2}=\left\langle
P^{2}\right\rangle _{\mathcal{\tilde{\eta}}(t)}-\left\langle P\right\rangle
_{\mathcal{\tilde{\eta}}(t)}^{2}. \label{C04}%
\end{equation}
The next step is to evaluate the expectation values $\left\langle
X\right\rangle _{\mathcal{\tilde{\eta}}(t)}$ , $\left\langle X^{2}%
\right\rangle _{\mathcal{\tilde{\eta}}(t)},$ $\left\langle P\right\rangle
_{\mathcal{\tilde{\eta}}(t)}$ and $\left\langle P^{2}\right\rangle
_{\mathcal{\tilde{\eta}}(t)}$ in the states $\psi(t)$ of $H(t)$ defined in
Eq. (\ref{schrod}). Using the $\mathcal{\tilde{\eta}}(t)$-inner product, and
after some calculations, we find that
\begin{align}
	\left\langle X\right\rangle _{\mathcal{\tilde{\eta}}(t)}  =\left\langle
	\psi(t)\right\vert F^{+}\mathcal{\eta}FX\left\vert \psi(t)\right\rangle
	=\left\langle \varphi\right\vert \rho F\tilde{\rho}^{-1}(t)x\tilde{\rho
	}(t)F^{+}\rho^{-1}\left\vert \varphi\right\rangle =\left\langle \varphi\right\vert
	x\left\vert \varphi\right\rangle ,\label{C05}%
\end{align}
\begin{align}
	\left\langle X^{2}\right\rangle _{\mathcal{\tilde{\eta}}(t)} =\left\langle
	\psi(t)\right\vert F^{+}\mathcal{\eta}FX^{2}\left\vert \psi(t)\right\rangle
	=\left\langle \varphi\right\vert \rho F\tilde{\rho}^{-1}(t)x^{2}\tilde{\rho
	}(t)F^{+}\rho^{-1}\left\vert \varphi\right\rangle =\left\langle \varphi\right\vert
	x^{2}\left\vert \varphi\right\rangle ,\label{C06}
\end{align}
\begin{align}
	\left\langle P\right\rangle _{\mathcal{\tilde{\eta}}(t)} =\left\langle
	\psi(t)\right\vert F^{+}\mathcal{\eta}FP\left\vert \psi(t)\right\rangle
	=\left\langle \varphi\right\vert \rho F\tilde{\rho} ^{-1}(t)p\tilde{\rho
	}(t)F^{+}\rho^{-1}\left\vert \varphi\right\rangle =\left\langle \varphi\right\vert
	p\left\vert \varphi\right\rangle ,\label{C007}
\end{align}
\begin{align}
	\left\langle P^{2}\right\rangle _{\mathcal{\tilde{\eta}}(t)}  =\left\langle
	\psi(t)\right\vert F^{+}\mathcal{\eta}FP^{2}\left\vert \psi(t) \right\rangle
	=\left\langle \varphi\right\vert \rho F\tilde{\rho}^{-1}(t)p^{2}\tilde{\rho
	}(t)F^{+}\rho^{-1}\left\vert \varphi\right\rangle =\left\langle \varphi\right\vert
	p^{2}\left\vert \varphi\right\rangle ,\label{C008}
\end{align}
indeed
\begin{equation}
\left(  \Delta X\right)  _{\mathcal{\tilde{\eta}}(t)}\left(  \Delta P\right)
_{\mathcal{\tilde{\eta}}(t)}=\left(  \Delta x\right)  \left(  \Delta p\right)
\geq\frac{1}{2}\left\vert \left\langle \left[  x,p\right]  \right\rangle
\right\vert .
\end{equation}
Thus, the uncertainty relation remains invariant. This equivalence ensures
that the uncertainty relation derived within the Hermitian framework is
preserved in the pseudo-Hermitian formulation.

\section{Particle in a complex TD linear potential}

The application concerns a class of one dimensional model of a particle with
TD mass $m(t)=m_{0}\lambda(t)$ subjected to the action of the following
complex TD linear potential $V(x,t)=$ $i\sqrt{\lambda(t)}x$. The corresponding
class of Hamiltonians is of the form
\begin{equation}
H(t)=\frac{p^{2}}{2m_{0}\lambda(t)}+i\sqrt{\lambda(t)}x, \label{B3}%
\end{equation}
with $m_{0}$ is a characteristic parameter of the system, the TD function $\lambda(t)$ is a 
strictly positive real function ($\lambda(t)\neq 0$) that can be choosen to describe a specific 
quantum system. 

To find the exact solution of the explicitly TD Schr\"{o}dinger
equation (\ref{schrod}), we use a suitable unitary transformation $F(t)$ defined as
\cite{fan1,kecita}
\begin{equation}
F(t)=\exp\left[ i\frac{m_{0}\dot{\lambda}(t)}{4\lambda(t)}x^{2}\right]
\exp\left[ - \frac{i}{2}\left\{  x,p\right\}  \ln\left(  \sqrt{\lambda
(t)}\right)  \right] , \label{B03}%
\end{equation}
which transforms the canonical operators $x$ and $p$ and their squares
according to
\begin{equation}
FxF^{+}=\frac{x}{\sqrt{\lambda(t)}},\ \ \ \ FpF^{+}=p\sqrt{\lambda(t)}%
-\frac{m_{0}\dot{\lambda}(t)}{2\sqrt{\lambda(t)}}x\ , \label{B04}%
\end{equation}%
\begin{equation}
Fp^{2}F^{+}=\lambda(t)p^{2}-\frac{1}{2}m_{0}\dot{\lambda}(t)\left\{
x,p\right\}  +\frac{m_{0}^{2}\dot{\lambda}^{2}(t)}{4\lambda(t)}x^{2},\text{
\ \ }Fx^{2}F^{+}=\frac{x^{2}}{\lambda(t)}.\text{\ } \label{B05}%
\end{equation}
Using Eqs. (\ref{B03}), (\ref{B04}) and (\ref{B05}) and after basic but
tedious calculations, the transformed Hamiltonian\ (\ref{B3}) becomes
\begin{equation}
\mathcal{H}(t)=\frac{p^{2}}{2m_{0}}+\frac{1}{2}m_{0}{\Omega}^{2}(t)x^{2}+ix,
\label{B06}%
\end{equation}
with
\begin{equation}
{\Omega}^{2}(t){=}\left(  \frac{1}{4}\frac{\dot{\lambda}^{2}(t)}{\lambda
^{2}(t)}-\frac{\ddot{\lambda}(t)}{2\lambda(t)}\right) . \label{B07}%
\end{equation}
The goal is to render the Hamiltonian (\ref{B06}) time-independent. To achieve
this, ${\Omega}^{2}(t)$ must equal to a constant denoted ${\Omega}_{0}^{2}$,
which leads to the following auxiliary equation
\begin{equation}
\ddot{\lambda}(t)-\frac{\dot{\lambda}^{2}(t)}{2\lambda(t)}+2\lambda(t){\Omega
}_{0}^{2}=0. \label{B08}%
\end{equation}
By introducing the change $\lambda(t)=\frac{1}{\alpha^{2}(t)}$, Eq.
(\ref{B08}) reduces to the following simple form
\begin{equation}
\ddot{\alpha}(t)+{\Omega}_{0}^{2}\alpha(t)=0. \label{B09}%
\end{equation}
For ${\Omega}_{0}^{2}>0$, we show that the Hermitian Hamiltonian (\ref{similarity}) describes a harmonic oscillator,  while for
${\Omega}_{0}^{2}<0$ an inverted oscillator. Both cases will be examined below.

\subsection{The case of a harmonic oscillator}

For $\Omega_{0}^{2}\mathcal{>} 0$ with  $\Omega_{0}\mathcal{>} 0$, the solution of Eq. (\ref{B09}) is 
\begin{equation}
	\alpha(t)=A_{1} \, e^{{i\Omega}_{0} t}+B_{1} \, e^{-i{\Omega}_{0} t} , \label{alpha1}
\end{equation}
then 
\begin{equation}
\lambda(t)=\left( A_{1} \, e^{{i\Omega}_{0} t}+B_{1} \, e^{-i{\Omega}_{0} t} \right)^{-2}, \label{lambda1}
\end{equation}
and the corresponding expression of the Hamiltonian (\ref{B3}) is    
\begin{equation}
	H(t)=\left( A_{1} \, e^{{i\Omega}_{0} t}+B_{1} \, e^{-i{\Omega}_{0} t} \right)^{2}\frac{p^2}{2m_{0}} + i \left( A_{1} \, e^{{i\Omega}_{0} t}+B_{1} \, e^{-i{\Omega}_{0} t} \right)^{-1} x. \label{BB3}%
\end{equation}
Therefore, the resulting time-independent non-Hermitian Hamiltonian
\textrm{\ }
\begin{equation}
\mathcal{H}_{0}^{PH}=\frac{p^{2}}{2m_{0}}+\frac{1}{2}m_{0}\Omega_{0}^{2}%
x^{2}+ix, \label{pseudo}
\end{equation}
is\textrm{\ }$\eta$-pseudo-Hermitian such that $\alpha(t)$ satisfies the
auxiliary equation (\ref{B09}), where $\eta$ is
\begin{equation}
\eta=\exp\left[ \frac{2p}{m_{0}\Omega_{0}^{2}}\right] .
\end{equation}

Moreover, $\mathcal{H}_{0}^{PH}$ can be related to the hermitian Hamiltonian
$h$ of the standard harmonic oscillator, via the similarity transformation
(\ref{similarity}), as
\begin{equation}
h\;=\;\rho\,\mathcal{H}_{0}^{PH}\,\rho^{-1}\;=\;\frac{p^{2}}{2m_{0}}+\frac
{1}{2}\,m_{0}\Omega_{0}^{2}\,x^{2}\;+\;\frac{1}{2m_{0}\Omega_{0}^{2}},
\end{equation}
where $\rho$ is defined as
\begin{equation} 
\rho=\sqrt{\eta}=\exp\left[  \frac{p}{m_{0}\Omega_{0}^{2}}\right]  ,
\end{equation}
and the eigenstates $\{|\varphi_{n}\rangle\}$ of $h$ are (where $\hbar$ is recovered)
\begin{equation}
	\varphi_{n}(x)=N_{1}\,exp\left[-\,\frac{m_{0}\Omega_{0}}{2\hbar}x^{2}\right]
H_{n}\!\left(x\sqrt{\tfrac{m_{0}\Omega_{0}}{\hbar}}\right), \label{wave}%
\end{equation}
with the eigenvalue $\;E_{n}=\hbar\Omega_{0}(n+\tfrac{1}{2})+\tfrac{1}{2m_{0}\Omega_{0}^{2}}$.

Under this construction, the solution (\ref{gsol}) associated with the Hamiltonian (\ref{BB3}) is
\begin{equation}
\lvert\psi(t)\rangle=F^{-1}(t)\,e^{-{iE_{n}t/}{\hbar}}\,\rho^{-1}\lvert\varphi
_{n}\rangle,\label{sol1}%
\end{equation}
and the probability density computed using the $\tilde{\eta}(t)$-inner
product
\begin{equation}
\bigl\lvert\psi (t)\bigr\rvert_{\tilde{\eta}}^{2}=\bigl\lvert\varphi
_{n}(x)\bigr\rvert^{2},
\end{equation}
is time-independent and can be normalized. 

Furthermore, the exact analytical solution (\ref{sol1})
\begin{align}
\left\vert \psi(t)\right\rangle    =F^{-1}\left\vert \phi(t)\right\rangle 
=\exp\left[  +\frac{i}{2}\left\{  x,p\right\}  \ln\left(  \sqrt{\lambda
(t)}\right)  \right]  \exp\left[  -i\frac{m_{0}\dot{\lambda}(t)}{4\lambda
(t)}x^{2}\right]  \left\vert \phi(t)\right\rangle,\label{AA2}%
\end{align}
represents, in the position representation, a squeezed state in terms of the Hermite polynomials with complex coefficients, as follows
\begin{equation}
\psi(x,t)=\exp\left[  -i\frac{m_{0}\dot{\lambda}(t)}{4\lambda^{2}(t)} x^{2}\right]   \lambda^{1/4}(t) \left\langle
x\sqrt{\lambda(t)} \mid\phi(t)\right\rangle, \label{AAA2} 
\end{equation}
where   
\begin{align}
	\left\langle x\sqrt{\lambda(t)}\mid\phi(t)\right\rangle  &
	=\left\langle x\sqrt{\lambda(t)}\right\vert e^{- {iE_{n}t/}{\hbar}}\,\rho^{-1}%
	\lvert\varphi_{n}\rangle= e^{- {iE_{n}t/}{\hbar}}\, \varphi_{n}\left(  x\sqrt{\lambda(t)}%
	+\frac{i\hbar}{m_{0}\Omega_{0}^{2}}\right)  \nonumber\\
	&  =\left[  \frac{\sqrt{m_{0}\Omega_{0}}}{n!2^{n}\sqrt{\pi\hbar}}\right]
	^{1/2} e^{- {iE_{n}t/}{\hbar}}\,  \exp\left[-\frac{m_{0}\Omega_{0}}{2\hbar
	}\left( x\sqrt{\lambda(t)} +\frac{i\hbar}{m_{0}\Omega_{0}^{2}}\right)^{2}\right] \times \nonumber\\ 
	& \qquad H_{n}\left[ \sqrt{\tfrac{m_{0}\Omega_{0}}{\hbar}} \left( x\sqrt{\lambda(t)} + \frac{i\hbar}{m_{0} \Omega_{0}^{2}}\right)  \right],
\end{align} 
and $\lambda(t)$ is given by Eq.(\ref{lambda1}). We note that the following formula was used in Eq. (\ref{AAA2})
\begin{equation}
	\left\langle x\right\vert \exp\left[  +\frac{i}{2}\left\{  x,p\right\}  \ln\left(  \sqrt{\lambda
		(t)}\right)  \right] =\exp\left[ \frac{1}{2} \ln\left(\sqrt{\lambda(t)}\right)  \right] \left\langle 
	x\sqrt{\lambda(t)} \right\vert.\label{A35'}%
\end{equation}

\subsubsection*{Uncertainty Relation}

Using the $\tilde{\eta}(t)$-inner product and after straightforward algebra,
the first and second moments of $X$ and $P$ are 
\begin{equation}
\bigl\langle X\bigr\rangle_{\tilde{\eta}}=\bigl\langle P\bigr\rangle_{\tilde
{\eta}}=0,\quad\bigl\langle X^{2}\bigr\rangle_{\tilde{\eta}}=\frac{\hbar
}{m_{0}\Omega_{0}}\bigl(n+\tfrac{1}{2}\bigr),\quad\bigl\langle P^{2}%
\bigr\rangle_{\tilde{\eta}}= \hbar m_{0}\Omega_{0}\bigl(n+\tfrac{1}{2}\bigr).
\end{equation}
Hence, the variances are
\begin{equation}
(\Delta X)_{\tilde{\eta}}=\sqrt{\langle X^{2}\rangle_{\tilde{\eta}}}%
=\sqrt{\frac{\hbar}{m_{0}\Omega_{0}}\Bigl(n+\tfrac{1}{2}\Bigr)},\quad(\Delta
P)_{\tilde{\eta}}=\sqrt{\langle P^{2}\rangle_{\tilde{\eta}}}=\sqrt
{\hbar m_{0}\Omega_{0}\Bigl(n+\tfrac{1}{2}\Bigr)}.
\end{equation}
Then we obtain
\begin{equation}
	(\Delta X)_{\tilde{\eta}}\,(\Delta P)_{\tilde{\eta}}\;= \hbar \Bigl(n+\tfrac
	{1}{2}\Bigr), \label{eq:finalUP}%
\end{equation}
and using $[X,P]=i\hbar$, the Heisenberg uncertainty relation (\ref{C01}) is
\begin{equation}
(\Delta X)_{\tilde{\eta}}\,(\Delta P)_{\tilde{\eta}}\;\geq\;\tfrac{\hbar}{2},\label{heis}%
\end{equation}
which is manifestly real, greater than or equal to $\hbar/2$ and coincides exactly with
that of the standard Hermitian harmonic oscillator.

\subsection{The case of the inverted oscillator}
For ${\Omega}_{0}^{2}=-{\omega}_{0}^{2}\mathcal{<}0$ with  $\omega_{0}\mathcal{>} 0$, the solution of Eq. (\ref{B09}) is
\begin{equation}
	\alpha(t)=A_{2} \, e^{{\omega}_{0} t}+B_{2} \, e^{-{\omega}_{0} t},  \label{alpha2}
\end{equation}
then 
\begin{equation}
	\lambda(t) = \big( A_{2} \, e^{{\omega}_{0} t} + B_{2} \, e^{-{\omega}_{0} t} \big)^{-2}, \label{lambda2}
\end{equation}
and the corresponding Hamiltonian (\ref{B3}) is   
\begin{equation}
	H(t)= \big( A_{2} \, e^{{\omega}_{0} t}+B_{2} \, e^{-{\omega}_{0} t} \big)^{2}\frac{p^2}{2m_{0}}  + i \big( A_{2} \, e^{{\omega}_{0} t}+B_{2} \, e^{-{\omega}_{0} t} \big)^{-1} x, \label{BB33}%
\end{equation}
 and the time-independent non-Hermitian Hamiltonian (\ref{pseudo})
\begin{equation}
	\mathcal{H}_{0}^{PH}=\frac{p^{2}}{2m_{0}}-\frac{1}{2}m_{0}\omega_{0}^{2}x^{2}+ix,
\end{equation}
is\textrm{\ }$\eta$-pseudo-Hermitian provided that $\alpha(t)$ satisfies the
auxiliary equation (\ref{B09}), where $\eta_{i}$ is
\begin{equation}
	\eta_{i}=\exp\left[- \frac{2p}{m_{0}\omega_{0}^{2}}\right].
\end{equation}
Further, $\mathcal{H}_{0}^{PH}$ is related to the hermitian Hamiltonian
$h$ of the inverted oscillator, via the similarity transformation
(\ref{similarity}), as
\begin{equation}
	h\;=\;\rho_{i}\,\mathcal{H}_{0}^{PH}\,\rho_{i}^{-1}\;=\;\frac{p^{2}}{2m_{0}}-\frac
	{1}{2}\,m_{0}\omega_{0}^{2}\,x^{2}\;-\;\frac{1}{2m_{0}\omega_{0}^{2}},
\end{equation}
which is unbounded from below, and its eigenstates $\{|\varphi_{n}^{i}\rangle\}$ are \cite{chrus,maamache4,bagarello22}
\begin{equation}
	\varphi_{n}^{i}(x)=N_{2}^{\pm}\,exp\left[\mp \,\frac{i m_{0}\omega_{0}}{2\hbar}x^{2}\right]
	H_{n}\!\left(x e^{\pm i\pi/4}\sqrt{\tfrac{m_{0}\omega_{0}}{\hbar}}\right), \label{wave2}
\end{equation}
and can also be expressed in terms of parabolic cylindrical functions \cite{chrus,subram,ullinger,sundaram}. 

The solution (\ref{wave2}) are not square integrable, the eigenvalues are continuous and doubly degenerate $(E,-E)$, with $ -\infty < E < +\infty$, and instability occurs because the spectrum is not bounded from below \cite{sundaram}. However, although $h$ is self-adjoint, it should be noted that the inverted oscillator exhibits a purely imaginary spectrum \cite{chrus,maamache4,bagarello22,subram,ullinger,sundaram} when the PT-symmetry is broken \cite{sundaram}.

The exact analytical solution corresponding to the Hamiltonian (\ref{BB33}) takes the form, in the position representation, 
of a squeezed state in terms of the Hermite polynomials with complex coefficients, as
\begin{equation}
	\psi(x,t)=\exp\left[  -i\frac{m_{0}\dot{\lambda}(t)}{4\lambda^{2}(t)} x^{2}\right]   \lambda^{1/4}(t) \left\langle
	x\sqrt{\lambda(t)} \mid\phi(t)\right\rangle, \label{AAA22} 
\end{equation}
where   
\begin{align}
	\left\langle x\sqrt{\lambda(t)}\mid\phi(t)\right\rangle  &
	=\left\langle x\sqrt{\lambda(t)}\right\vert e^{- {iE t/}{\hbar}}\,\rho^{-1}%
	\lvert\varphi_{n}^{i}\rangle= e^{- {iE t/}{\hbar}}\, \varphi_{n}^{i} \left(  x\sqrt{\lambda(t)}%
	-\frac{i\hbar}{m_{0}\omega_{0}^{2}}\right)  \nonumber\\
	&  = N_{2}^{\pm} \, e^{- {iE t/}{\hbar}}\,  \exp\left[\mp \frac{i m_{0}\omega_{0}}{2\hbar
	}\left( x\sqrt{\lambda(t)} - \frac{i\hbar}{m_{0}\omega_{0}^{2}}\right)^{2}\right] \times \nonumber\\ 
	& \qquad H_{n}\left[e^{\pm i\pi/4} \sqrt{\tfrac{m_{0}\omega_{0}}{\hbar}} \left( x\sqrt{\lambda(t)} - \frac{i\hbar}{m_{0} \omega_{0}^{2}}\right)  \right],
\end{align} 
and $\lambda(t)$ is given by Eq. (\ref{lambda2}).

\section{Conclusion}

This work provides a coherent framework for the treatment of TD 
non-Hermitian systems, showing how TD unitary transformations can
simplify the study while preserving physical properties. Indeed, in order to solve the Schrödinger equation for a class of TD non-Hermitian  Hamiltonians $H(t)$ convertible into a time-independent pseudo-Hermitian Hamiltonian $\mathcal{H}_{0}^{PH}$ via a unitary transformation $F(t)$, and $\mathcal{H}_{0}^{PH}$ can be related to a Hermitian Hamiltonian $h$ via a similarity transformation. The problem
therefore reduces to solve the Schr\"{o}dinger equation for the Hermitian
Hamiltonian $h$, and thus the general solution of the original system can be
deduced. We constructed the appropriate $\tilde{\eta}(t)$-inner product for
the Hilbert space associated with $H(t)$, ensuring that the norm of the
quantum state remains preserved over time evolution, and defined the relevant
uncertainty relation for position and momentum using the same inner product. 

As an application, we investigated the solution of the Schrodinger equation of a class of 
Hamiltonians for a particle with time-dependent mass influenced by a complex linear
time-dependent potential. For this model, two time-independent pseudo-Hermitian Hamiltonians $\mathcal{H}_{0}^{PH}$ have been derived, and their corresponding Hermitian Hamiltonians $h$ being those of the harmonic oscillator and the inverted oscillator respectively. The two exact analytical
solutions have been obtained in terms of the Hermite polynomials with complex coefficients. When the Hermitian Hamiltonian is the harmonic oscillator, and using the $\tilde{\eta}(t)$-inner product, we have shown that the uncertainty relation for the position and momentum is always real and greater than or equal to $\hbar/2$, i.e., it is physically consistent.

\section*{Acknowledgements} 
This work is supported by the Algerian Ministry of Higher Education and Scientific Research under 
the PRFU Project Number: B00L02UN180120210003.
{\tiny }

\end{document}